\documentclass[amsmath,amssymb,twocolumn,aps,floatfix,showpacs]{revtex4}
\usepackage[dvips]{epsfig}

\begin{document} 

\title{Helical Tubes in Crowded Environments}
\author{Yehuda Snir}
\author{Randall D. Kamien}
\affiliation{Department of Physics and Astronomy, University of Pennsylvania, Philadelphia, PA 19104-6396}
\begin{abstract}
When placed in a crowded environment, a semi-flexible tube is forced to fold so as to make a more compact shape. One compact shape that often arises in nature is the tight helix, especially when the tube thickness is of comparable size to the tube length. In this paper we use an excluded volume effect to model the effects of crowding. This gives us a measure of compactness for configurations of the tube, which we use to look at structures of the semi-flexible tube that minimize the excluded volume.  We focus most of our attention on the helix and which helical geometries are most compact. We found that helices of specific pitch to radius ratio 2.512 to be optimally compact. This is the same geometry that minimizes the global curvature of the curve defining the tube.  We further investigate the effects of adding a bending energy or multiple tubes to begin to explore the more complete space of possible geometries a tube could form. 
\end{abstract}
\date{\today}
\pacs{81.16.Fg, 02.10.Kn,02.40.-k,05.20.-y}
\maketitle

\section{introduction}

The universe is a sparse and lonely place, with vast amounts of empty space.  Most of us prefer to be in more dense regions, however, and it is precisely in these crowded environments that life occurs.  These dense regions raise the question of how to pack things efficiently.  Perhaps the most storied example is Kepler's conjecture that the face centered cubic (fcc) lattice is the densest way to
pack identical spheres. Though there was never much doubt that Kepler was right, a rigorous proof
defied the centuries until Hales \cite{Hales} developed a program of proof which established this
long sought result.  Free volume theory \cite{Kirkwood}, for instance, uses the density of different
lattices to establish their purely entropic free energy differences.  As the fcc lattice is the most densely packed, it allows for the greatest amount of translational entropy at any fixed volume fraction.
When considering infinite volume systems, it is straightforward to pose optimality questions in terms
of infinite periodic structures, whether it is the packing problem, the channel coding problem, the covering problem, {\sl etc.} \cite{Sloane}.  However, when considering finite systems, or clusters, 
optimality can be more subtle.  For instance, clusters of colloidal microspheres repeatably assemble into regular clusters that, in general, defy characterization through any packing principle \cite{Pine,Brenner}.
Here we consider the optimal packing of a semi-flexible tube.  Since the only degree of freedom for a single tube is its conformation, density is no longer a useful notion, as the tube's volume remains fixed.  In this paper we consider the conformation of a solid, semi-flexible tube which maximizes the free volume for a hard sphere interacting only through its excluded volume.

Other approaches have been proposed in defining the compact configurations of tubes.  For instance, when studying knots made of solid tubes of fixed radius (in $\mathbb{R}^3$), one can ask what geometry minimizes the length of the knot, {\sl i.e.} what is the tightest knot?  Though easily posed, this is a difficult question for which there is still no known minimizer for the simplest of knots, the trefoil \cite{KusSul}.  One approach is to minimize the {\sl global radius of curvature} \cite{Maddocks}, which we will discuss in detail in the following section.  Another approach is  to minimize the convex hull of the tube's conformation.  While both of these work for closed loops and knots, they are both problematic when
considering finite, open tubes.  In this case the straight tube is always optimal and so neither
captures the notion of a compact conformation.

Our approach, on the other hand, hinges on the Asakura-Oosawa depletion interaction \cite{AO}.  We immerse our tube in a bath of hard spheres and maximize their entropy or, equivalently, their
free volume.  This gives conformations of the tube that capture the notion of compactness.  The structure of this paper will be as follows.  In section II we describe depletion volume theory, first for binary colloidal systems, and extend it to other systems. In section III we describe how to minimize the excluded volume of a free tube of finite thickness.  We find that helices can arise for a range of tube lengths and support this with numerical integration of the free volume for spheres of varying radii.  In the limit that the sphere radius is much smaller than the tube thickness, we are able to evaluate the free volume analytically and argue that this limit should generate those
configurations found by minimizing the global radius of curvature for closed knots.  In section IV we consider the effect of additional interactions and study tube rigidity and tube-tube interactions.  Section V summarizes our results. 
\begin{figure}[t]
\begin{center}
\includegraphics[width=3in]{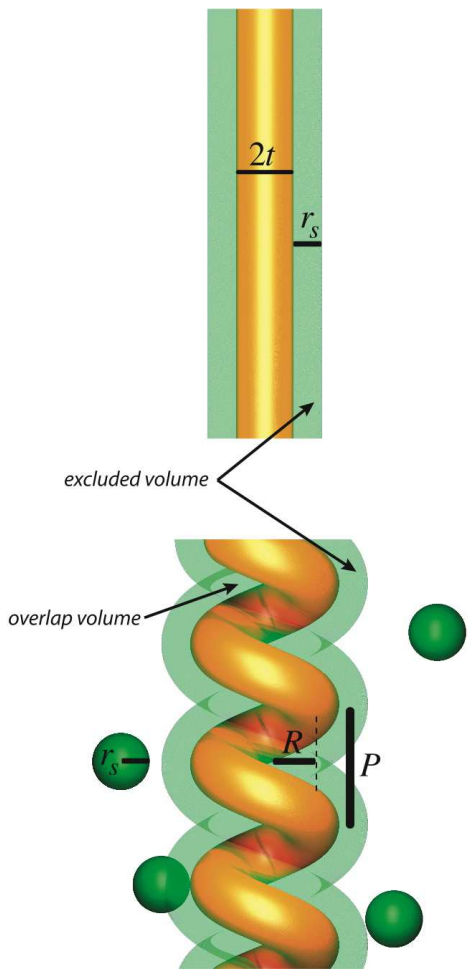} 
\caption{{Excluded volume (green) for a straight and bent tube where the spheres cannot be. When the tube bends into the helical conformation there is overlap of the excluded volume between the layers of the helix and in the central core region of the helix. The excluded volume decreases by the overlap volume when the tube forms the helix.}}
\label{helixpics}
\end{center}
\end{figure}

\begin{figure}[b]
\begin{center}
\includegraphics[width=3in]{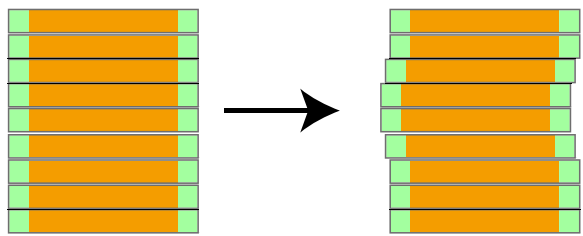} 
\caption{{If there is a section of tube where the excluded volume does not overlap with some other
region of excluded volume then a deformation of the tube can increase the overlap.  The move does not
disturb the remainder of the chain, just as the center of a stack of coins can be shifted without moving the other coins. }}
\label{pennyshift}
\end{center}
\end{figure}

\section{Entropy and Depletion Interactions}
The depletion model of Asakura and Oosawa, first developed to study polymer solutions, relies on the virial expansion for a mixture of large and small spheres \cite{AO}.   Consider two large hard spheres of radius $r_l$ in a solution of smaller hard spheres of radius $r_s$.  The centers of the two large spheres can get no closer than $2r_l$ apart while the small spheres can get no closer than $r_l+r_s$ to the large spheres before they overlap.  Thus we can enshroud each large sphere with a halo of thickness $r_s$ which excludes the small spheres.  As the large spheres come closer, these excluded volume halos overlap and the free volume ($V_{free}$) for the small spheres increases so that for $N_s$ small spheres, the free energy changes by $\Delta F=-k_BTN_s \Delta\ln(V_{free}/V_0)\approx -k_BTN_s \Delta V_{free}/V_{free}=-k_BTn V_{overlap}$ where $V_0$ is a characteristic volume for the small colloids, $n=N_s/V_{free}$ is the concentration of small spheres, and $V_{overlap}=\Delta V_{free}$ is the overlap of the excluded volumes which is small compared to the free volume.  This results in an effective short range attraction between the large spheres at distances of order $r_s$, an effect which has been seen in binary colloid systems \cite{binColloid} and colloid-polymer systems \cite{Crocker}. The free energy comes from purely geometric considerations: what is the arrangement of large spheres that maximizes the overlap of their halos and how large is that overlap volume? For two spheres the answer is obvious, the largest overlap occurs when the two spheres touch, and the overlap volume in this case is approximately $2 \pi r_l r_s^2$ when $r_s   \ll r_l$.   This result can be corroborated in detail through the virial expansion, for instance.  

The standard geometrical picture allows us to generalize, without developing the machinery of the virial expansion, to other systems that interact via a hard-core interaction. For instance, a large sphere is attracted to a wall since its excluded volume overlaps with the wall's excluded volume. Thus the presence of boundaries complicates the problem of finding the configuration of large spheres which maximizes the overlap volume. Since the overlap of a sphere's and flat wall's excluded volume is roughly twice that for two spheres, approximately $4 \pi r_l r_s^2$, this attraction is large enough to drive two-dimensional crystallization along walls in place of three dimensional crystallization in bulk \cite{binColloid}. In this paper we replace the large spheres by a hard (but flexible) tube of radius $t$.  In analogy to the halo around the large spheres there is a halo of excluded volume around the tube for all points within $r_s$ of the tube. In the previous example the large spheres had an effective attraction, while in our situation the tube has an effective self attraction forcing it to bend. When the tube bends it is able to overlap the excluded volume from the two tube segments, leading to a decrease in the overall free energy.  Thus we seek the shape of the tube which minimizes the free energy of the small spheres.  We may also consider the optimal shapes of closed tubes and tubes
tied into knots.  We will discuss this in the following.

\section{Free Tubes}
The space of possible configurations of the tube is difficult to characterize, making it challenging to find the shape of the excluded volume (see, for instance, figure \ref{helixpics}). For an ``ideal'' tube (a tube that is perfectly hard, incompressible, but can bend flexibly) the configuration space is additionally constrained by the tube thickness.  Because the tube cannot bend and overlap itself, we can greatly reduce the configuration space by considering shorter tube segments.  At all segments of the tube the radius of curvature of the centerline must be larger than the tube radius, $t$, due to the tube thickness.   As the tube gets longer, it becomes necessary to consider contacts from parts of the tube
that are not ``nearby'' and then we use the global radius of curvature to account for this.   

\subsection{Global Radius of Curvature and the Small Sphere Limit}
The radius of curvature at a point $\bf x$ on a curve $\cal C$ is the radius of the circle that passes through $\bf x$ and two adjacent points (in the continuum this can be defined through the obvious limiting procedure).  Given three non-colinear points ${\bf x}_i$, $i=1,2,3$, we can construct
the radius of the circle that passes through all three:
\begin{equation}\label{cradius}
r({\bf x}_1,{\bf x}_2,{\bf x}_3) = {1\over 2}\sqrt{\left({v^2-uv + w^2\over  w}\right)^2 + u^2}
\end{equation}
with
\begin{eqnarray}
u&=&\vert {\bf x}_2-{\bf x}_1\vert\nonumber\\
v&=&{\left({\bf x}_3-{\bf x}_1\right)\cdot\left({\bf x}_2-{\bf x}_1\right)\over \vert{\bf x}_2-{\bf x}_1\vert}\nonumber\\
w&=&{\left\vert\left({\bf x}_3-{\bf x}_1\right)\times\left({\bf x}_2-{\bf x}_1\right)\right\vert\over \vert{\bf x}_2-{\bf x}_1\vert}
\end{eqnarray}
and where $w\neq 0$ since the three points do not lie on a line.  If we consider a continuous curve ${\bf R}(s)$, parameterized by arclength $s$ and choose ${\bf x}_1={\bf R}(s)$,  ${\bf x}_2={\bf R}(s+ds)$, and ${\bf x}_3={\bf R}(s-ds)$, then $u=ds +{1\over 2}\kappa ds^2$, $v=-ds/(1 +{1\over 2}\kappa ds)$ and $w=\kappa ds^2/(1+{1\over 2}\kappa ds)$, where $\kappa$ is
the curvature of the curve.  Then 
$\lim_{ds\rightarrow 0} r({\bf x}_1,{\bf x}_2,{\bf x}_3) = 1/\kappa$, recovering the radius of curvature.   Using the fact that the area of the triangle is $A({\bf x}_1,{\bf x}_2,{\bf x}_3) = {1\over 2}\vert \left({\bf x}_i-{\bf x}_j\right)\times\left({\bf x}_k-{\bf x}_j\right) \vert$ for $i,j$, and $k$ distinct, we can rewrite (\ref{cradius}) as 
\begin{equation}
r({\bf x}_1,{\bf x}_2,{\bf x}_3) = {\vert {\bf x}_1-{\bf x}_2\vert \vert{\bf x}_2 -{\bf x}_3\vert \vert {\bf x}_3-{\bf x}_1\vert\over 4 A({\bf x}_1,{\bf x}_2,{\bf x}_3)}.
\end{equation}

Gonzalez and Maddocks \cite{Maddocks}\  introduced  the global radius of curvature $r_g({\bf x})$ to take into account contacts from points far away on the curve:
\begin{equation}
r_g({\bf x})= \min_{{\bf y},{\bf z} \in {\mathcal C}} r({\bf x},{\bf y},{\bf z}).
\end{equation}
Just as the tube around $\cal C$ can be no thicker than the radius of curvature, $t\le 1/\kappa$, the global condition becomes 
\begin{equation} 
t\le  \Delta[{\mathcal C}]=\min_{{\bf x}\in{\mathcal C}} r_g({\bf x}).
\end{equation}
In \cite{Maddocks}\ it was shown that a closed curve of fixed length which maximizes the thickness $\Delta[{\mathcal C}]$ has a constant value of $r_g({\bf x})=r_g^*$ for all points ${\bf x}$
where it is curved and a value of $r_g({\bf x})\ge r_g^*$ when the curve is straight.  We seek instead a curve which maximizes the overlap volume $V_o[{\mathcal C},r_s/t]$ which
depends on both the curve ${\mathcal C}$ and possibly the ratio of the radius of the depleting spheres to the thickness of the tube.    We have found \cite{Brevia} that when $r_s/t \rightarrow 0$, the tubes take on conformations for which the thickness of the tube $t$ is precisely equal to the constant, global radius of curvature.  Though this may not
be surprising, it is certainly not obvious and does not hold in general -- as the depleting spheres grow, the optimal tube configuration changes.  
\begin{figure}[t]
\begin{center}
\includegraphics[width=3in]{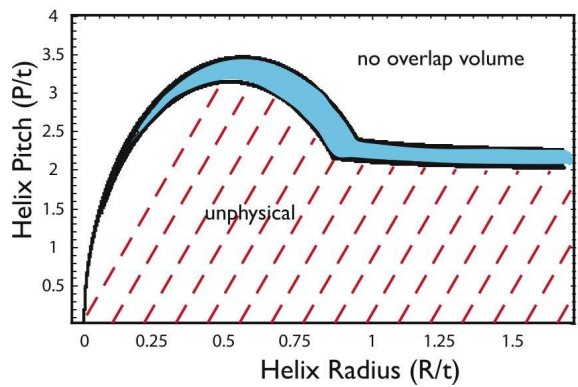} 
\caption{{Allowed helix pitch and radius for a hard tube. Helices with pitches and radii in the dashed region make unphysical configurations of the tube.  When $r_s=t/10$, only helices with pitches and radii in the shaded region have any overlap volume.  }}
\label{helixparam}
\end{center}
\end{figure}

\begin{figure}[t]
\begin{center}
\includegraphics[width=3in]{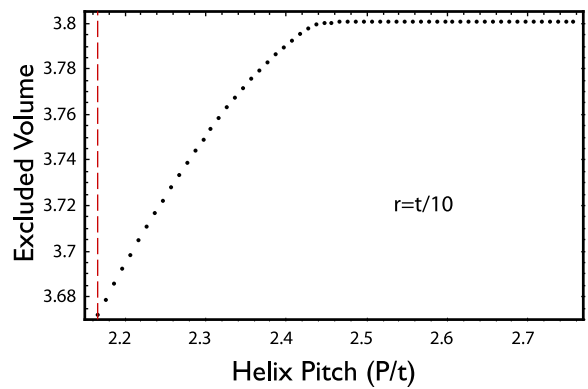} 
\caption{{Excluded volume as pitch increases above the minimum allowed value $P_{min}$ (dashed line). For small increases above $P_{min}$ the excluded volume increases linearly with slope 0.656 and plateaus as the turns separate enough so that there is no overlap of the excluded volume. We show our calculation for $r_s=t/10$.}}
\label{extraP}
\end{center}
\end{figure}

We argue that as $r_s/t\rightarrow 0$, if a smooth tube configuration exists which maximizes the overlap volume, then $r_g({\bf x})=t$ for all points on the curve.    To see this,
consider a point $\bf x$ for which $r_g({\bf x})\ne t$: since the local radius of curvature cannot be smaller than $t$ for a solid tube, $r_g$ cannot be
smaller than $t$.   Thus
we consider the case that $r_g({\bf x})=\rho>t$.  By assumption, the curve is smooth and so $r_g>\rho-\epsilon>t$ in a neighborhood of ${\bf x}$.  Thus there is an $r_s$ small enough so that 
there is no overlap of the excluded volume halo near ${\bf x}$.  Keeping $r_s$ fixed, we can move the curve in at ${\bf x}$ to achieve some overlap {\sl without} reducing the overlap with some other part of the tube {\sl since there was no overlap at the outset.}   We depict this ``stack of penny shift'' in Figure \ref{pennyshift}.   Thus, we can increase the overlap volume if $r_g({\bf x})>t$ and so that configuration cannot have the maximal overlap volume.  Hence for small depleting spheres the ideal tube shape will have $r_g=t$ along the whole curve.  We know no argument to show that this is a unique configuration, and it is entirely possible that there are multiple configurations for which $r_g=t$, only some (or one) of which maximizes the overlap volume.  Note that this discussion required arbitrarily small depleting spheres.  As the spheres grow, they begin to ``see the forest for the trees'' and, for instance, prefer straighter segments of
tube to increase the local overlap volume  -- surfaces which bend away from the sphere have a smaller overlap volume than surfaces which bend towards
the spheres.  Thus, our approach allows us to both see the effect of finite depletor size and the optimal geometry of Gonzalez and Maddocks.

\begin{figure}[b]
\begin{center}
\includegraphics[width=3in]{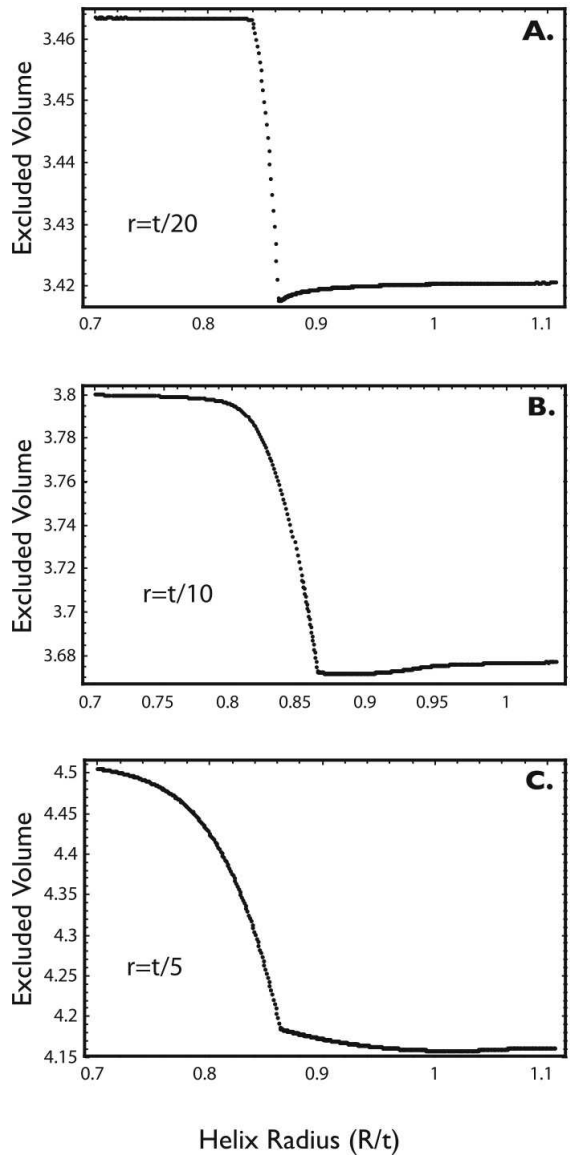} 
\caption{{Excluded volume as helix radius is changed for various sphere sizes. At small helix radius the geometry is constrained by the local curvature and the the layers of the helix do not touch. The big drop in excluded volume occurs when the helix turns start to have overlap between them. For small spheres $r_s=t/20$ (A.) the minimum excluded volume occurs at the maximally compact helix with radius 0.86218. As the sphere sizes grow the minimum excluded volume occurs at larger radius, for spheres of radius $r_s=t/10$ the minimum occurs at 0.88554 (B.) and at 1.01875 for spheres of radius $r_s=t/5$ (C.).}}
\label{freetube}
\end{center}
\end{figure}

\subsection{Computation of the Excluded Volume}
We start by restricting our study to helices as they are curves of constant curvature which can be chosen so that $t=r_g$.  
Even when restricting our considerations to helices, there is a two parameter space of possible helix geometries.  Aside from their length $L$, all helices can be described by their pitch $P$ and radius $R$ (see figure \ref{helixpics}). The curve defining the centerline of the  helix is  
\begin{equation} \label{helixeq}
{\bf R}(\xi)=(R \cos(\xi), R \sin(\xi), P \xi /2 \pi) 
\end{equation}
 with $\xi \in[0,2\pi n]$ where $n$ is the number of helical turns the tube makes, $n= {L \over \sqrt{P^2+(2\pi R)^2}}$. In order to calculate the overlap volume, we need to draw discs in the plane perpendicular to the curve's unit tangent $\bf T$.  A convenient basis for this plane is made by the normal $\bf N$ and binormal $\bf B$ unit vectors.  With our parameterization, we have:  
      
\begin{eqnarray} \label{helixvecs}
{\bf T}(\xi)&=&{2\pi \over \sqrt{\left(2\pi R\right)^2+P^2}} \left[-R\sin{\xi}, R \cos{\xi}, P/2\pi\right] \nonumber\\
{\bf N}(\xi)&=&\left[\cos{\xi}, \sin{\xi}, 0\right]\nonumber\\
{\bf B}(\xi)&=&{P \over \sqrt{\left(2 \pi R\right)^2+P^2}} \left[\sin{\xi}, -\cos{\xi}, 2\pi R/P\right]
\end{eqnarray}
An arbitrary point in the tube is at 
\begin{equation} \label{helixDef}
{\bf H}(\xi, \phi, s) = {\bf R}(\xi)+s \left[\cos{\phi}\; {\bf N}(\xi)+\sin{\phi}\;{\bf B}(\xi)\right]
\end{equation}
for $\phi\in[0,2\pi]$, $s\in[0,t]$, and $\xi\in[0,2\pi n]$.
Not all values of $P$ and $R$ are allowed \cite{Pieranski}\ when the tube is prevented from self-intersecting. The local curvature has an upper bound $\kappa= {R \over (P/2\pi)^2+R^2}\le{1 \over t}$, and the distance between successive turns also must be greater than $t$. These constraints define the lower boundary of the accessible region in Figure \ref{helixparam}.       
Among all helices, the helix with $P/R=2.512$ is the helix geometry where the thickness $\Delta[{\mathcal C}]$ is equal to the local curvature and to the distance between successive turns, making it a compact helix \cite{Pieranski} and the minimal excluded volume configuration when $r_s/t\rightarrow 0$.

Fortunately, there is an upper bound to the relevant accessible region as well.  If the helix is too rarefied, there will be no overlap of the excluded volume.  Independent of tube configuration, an ``ideal'' tube with no self intersections will have volume $V=\pi t^2 L$.   Similarly, if there is no overlap of the excluded volume, the total excluded volume will be $V_{\rm excluded} = \pi (t+r_s)^2 L$.  We need not consider configurations for which the global curvature exceeds $t+r_s$ since an ideal tube of this larger radius can adopt these shapes.  This consideration leads to the upper boundary of the region in Figure \ref{helixparam}.

In the region between the two curves, we know of no simple formula for the total excluded volume and we have resorted to numerical integration.  We take advantage of the screw symmetry of the helix
to reduce the dimensionality of the integration.  Recall that Pappus's theorem gives the volume of a surface of revolution as the area of the cross section multiplied by the length that its centroid traverses.    Consider the straight line which lies at the center of the helix (the $\hat z$-axis in our parameterization): 
the intersection of the helix with its excluded volume halo and any half plane beginning on this line has
an area independent of the direction of the plane ({\sl i.e.} the angle in the $xy$-plane).  The volume does not change if we slide each section up or down to be in registry with the cut through the $xz$-plane (for $x\ge 0$).  Then, by Pappus, the volume is just the area of the cross section multiplied by $2\pi r_{\rm centroid}$.   To determine the area we pixelate the $xz$-plane and check if each point can be represented by ${\bf H}$ with $s\le t + r_s$.  We can then calculate the area and the centroid.
To properly compare the geometries we calculate the excluded volume per unit length, and find the optimal helix for a given sphere size.  We find, not surprisingly, that  for any helix radius, $R$, the minimum pitch always resulted in the largest overlap volume.  This follows since increasing the pitch opens up the helix and allows gaps between successive turns of the tube.  In Figure \ref{extraP} we graph the overlap volume as we increase the pitch beyond its allowed minimum for fixed $R$ and show how the excluded volume decreases as the pitch decreases at a given radius.  For small increases in the pitch the excluded volume increases linearly with $P$ until successive turns no longer touch and then there is no longer an overlap region.  The radius that minimizes the excluded volume depends on sphere size and we will denote it as $R^*_r$ in the following.

\subsection{Free Tube Results}

As we probe the space of possible helix parameters, the overlap volume goes through various regions of high and low overlap. This is depicted in Figures \ref{freetube} A)-C) for sphere radii  $r_s=t/20$, $ t/10$, and $t/5$, respectively. In all of these calculations, we set the pitch to its minimum allowed value for each $R$ since this strictly minimizes the excluded volume. The largest amount of overlap occurs when layers of the helix lie one on top of the other. Thus, for small $R$ when the layers can not achieve this geometry, the overlap is small.  When the radius of the helix gets large enough, however, the local curvature does not prevent the turns of the helix from being near each other and the overlap volume greatly increases leading to the sharp decrease in excluded volume seen in the figures.  The primary difference between different helices in this regime is the overlap in the central core region.  This accounts for the rise in the excluded volume as $R/t$ grows.

\begin{figure}[b]
\begin{center}
\includegraphics[width=3in]{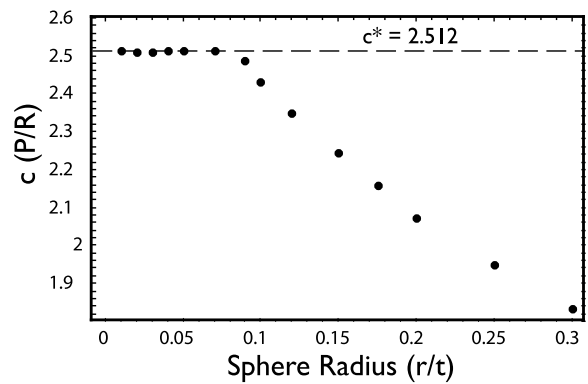} 
\caption{{Ratio of the optimal helix pitch to helix radius ($c=P/R$) for various sphere sizes. As the spheres get smaller, the value of $c$ approaches $c^*=2.512$ (dashed line). As the spheres get larger, c decreases because the helix radius is larger.}}
\label{helixPoverR}
\end{center}
\end{figure}

We note that as the sphere size grows the optimal value of $c=P/R$ decreases, though the general trend of excluded volume versus $R/t$ is unchanged. For instance, when  $r_s=t/20$ the minimum is almost exactly at the maximally compact helix with radius $R^*_{.05}=.86218 t$ and $P=2.1658 t$, or $c=P/R=2.512$. As the spheres get larger the optimal helix radius grows since the overlap of the excluded volume on the outside of the helix is {\sl increased} by having a flatter contact.  This effect competes with
the overlap in the core and can dominate the entropy.  When $r_s=t/10$, we find $R^*_{0.1} = 0.885544 t$ and $P=2.15444 t$, or $c=2.433$, a smaller value than that for the most compact tube \cite{Brevia}.  As shown in Figure \ref{helixPoverR}, $c$ continues to drop as $r_s$ increases.   

As noted in earlier work by Maritan {\sl et al.}, \cite{Maritan}, the optimal pitch to radius ratio is strikingly close to that for $\alpha$-helices in proteins.  This suggests that crowding itself \cite{Minton} can predict the geometry of proteins.  Of course, hydrogen bonds stabilize the $\alpha$-helix \cite{Pauling}, but our calculation provides plausible reason for why that geometry was preferred in the first place.
Our work is somehow dual to the use of site specific interactions which attempt to predict the secondary  structure of proteins based on sequence \cite{Dill,Zimm}.   It is in the spirit of Maritan {\sl et al.} \cite{Maritan, Maritan2} who find this particular helical geometry as a recurring motif for a solid tube with a variety of interactions which serve to confine it and force a compact structure. Indeed, their work has shown the helix and other protein structures as stable units which form purely on the
basis of geometry.   By constraining their study to tubes they were able to consider the intrinsic anisotropies of the system.  In the protein, the anisotropy of the chain of amino acids comes from the side groups and the carbon backbone -- both define the preferred direction \cite{tube}.  The helix can even arise from tethered hard spheres if the bond length between neighboring spheres is different than the hard sphere radius \cite{spheres}.

\subsection{Exact Analysis as $r_s/t\rightarrow 0$}

We can gain further insight into our numerics by studying the small sphere limit where $r_s\ll t$. In this limit, overlap of excluded volume amounts to two segments of the tube touching.   If the segments are not parallel their contact will only be at a point and the overlap volume will depend on their relative orientation. If the tube segments are aligned, however, then they will touch along a line.  This is the case in the helix with turns lying one on top of another.  Along a line of length $\ell$ the total overlap volume is the overlap area of two circles of radius $r+t$ a distance $2t$ apart times the $\ell$. This gives an overlap 
\begin{equation}\label{exV}
V_{overlap}={4 \sqrt{2} \over 3} r_s^{3/2} t^{1/2} \ell
\end{equation} 
We can also consider how this overlap volume changes when we pull the tubes apart to a distance $d<r_s$: 
\begin{eqnarray} \label{valpha}
V_{overlap}&=&{4 \sqrt{2} \over 3} r_s^{3/2} (1- \alpha)^{3/2} t^{1/2} \ell \nonumber\\
&=&{4\sqrt{2} \over 3}  r_s^{3/2}t^{1/2}\ell \big(1-{3 \over 2} \alpha+O(\alpha^2)\big)
\end{eqnarray}
where $\alpha=d/r_s$.  In Figure \ref{extraP} we have plotted the change in free volume per unit length as a function of the pitch $P$ for fixed $r_s$ and $R^*_{r_s}$.  This amounts to taking $d=P-P_{min}$ and we see that, indeed, the change in free volume is linear in $P$ and turns over when the tubes are no longer touching.  The slope is $m_{num}\approx 0.66$.   We can use (\ref{valpha}) to estimate the slope and find $m_{theory}\approx0.89\ell/L$.  What is the length $\ell$ of the line of contact?  By symmetry, this line is itself a helix with the same pitch as the tube but with a different radius.  Using the equation for the helix (\ref{helixeq}) we can consider where the piece of tube at $\xi_1=0$ (${\bf x}_1=[R,0,0]$) makes contact with another piece of tube centered at  ${\bf x}_2=[R \cos (\xi),R \sin (\xi),P\xi/2\pi]$.  The contact line is halfway between these two centers, at ${\bf x}_{mid} = {1 \over 2} \big[R(1+\cos(\xi)),R \sin(\xi),P\xi/2 \pi \big]$ with
a corresponding radius $\rho= R|\cos(\xi/2)|$ and helix length $\ell=n\sqrt{P^2+(2\pi R \cos(\xi/2))^2}$, where again $n$ is the number of turns in the helix.  In general, the
contact does not occur in a plane which contains the centerline of the helix -- only in the case of stacked tori or, equivalently, infinite $R$, does this happen.  Thus we find that $\ell/L<1$.  For fixed $R$ and $P$ we can find the value of $\xi$ which minimizes the distance $\vert {\bf x}_1-{\bf x}_2\vert $ and find
\begin{equation}\label{squiggle}
\frac{\sin\xi}{\xi} = -\frac{P^2}{(2\pi R)^2}
\end{equation}
Using our numerics for $P/R$ when $r_s=t/10$, we find roots at $\xi\approx\pm1.7\pi$ so that
$\ell/L\approx 0.91$.  This gives a final estimate of $m_{theory} \approx 0.81$, a reasonable, albeit sloppy, estimate for $m_{num}(\approx 0.66)$.  Further correction is possible by taking into account the reduction in overlap due to the local curvature. 

Armed with these analytic results, we can estimate at what length the helical conformation of the tube will give way to a larger superstructure ({\sl e.g.} pleated sheets).   For very short tubes, when $L<\pi t$, the tube can only form overlap of the excluded volume by bending into an arc of radius t.
Once the tube is long enough to make contacts from distant points, it will from a nascent helix, providing lines of contact instead of merely {\sl points} of contact.  The other possibility is that the tube bends back on itself, making
a pleat-like conformation.  
If we consider two helices next to each other, each of length $L/2$, we can determine whether the helix will bend back on itself by comparing the overlap volume of this configuration with that of a single helix.  Splitting the helix into two segments costs the contacts from one pair of turns, so the length
of contact goes down by $\ell_{\rm lost}>\sqrt{(2 \pi \rho)^2+P^2}$.  However, the pair of helices now gains a contact length $\ell_{\rm gain}< nP/2$.  Thus we expect the helix to be stable when
$nP/2>\sqrt{(2\pi\rho)^2+P^2}$ or, for the maximally compact helix with $P/R\approx 2.512$, we expect
stable helices when $n<4.3$.  It is interesting to note that the average length of $\alpha$-helices is 12 residues which corresponds to $n=3.3$, well within our simple estimates \cite{protein length}.
As the tube gets longer, the configuration that maximizes the excluded volume will surely get more complicated. In the limit of a very long tube, the problem becomes that of packing parallel tubes, an essentially two dimensional problem with the added cost of the hairpins necessary to bend the tube over. In this case the greatest overlap occurs when the tube bends back and forth to form an hexagonal lattice of cylinders similar to what is seen in hexagonally packed bundles of DNA. 

\subsection{On the Knot but Not the Knot:  Minimizing the Overlap Volume of a Trefoil}

\begin{figure}[t]
\begin{center}
\includegraphics[width=3in]{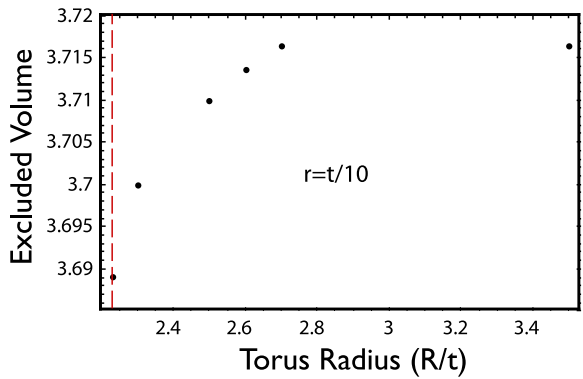} 
\caption{{Excluded volume versus torus radius for spheres of size $r=t/10$ and torus thickness $b=t/2$. The excluded volume is calculated for a tube forming a trefoil knot embedded on the surface of the torus defined by $R$ and $b$. The minimum excluded volume occurs at the minimum allowed radius, $R=1.1158$ (dashed line). }}
\label{trefoilgraph}
\end{center}
\end{figure}

We have argued that as $r_s/t\rightarrow 0$, the depletion interaction will drive the tube to its maximally compact conformation.  Indeed, our result for the optimal helix recapitulates the geometry which minimizes the tube length for a given thickness.  To further test our argument and make contact with the original motivation for developing the notion of global curvature, we will apply our analysis to the simplest knot, the trefoil.  As with the helix, we choose a family of trefoil knots and minimize over the free parameters.  There is a convenient class of trefoil knots, so called ``torus knots'', for which the centerline of the tube is embedded on a two-torus.  Though this obviously does not span the whole space of such knots, it is convenient because of its analytic form:   
\begin{eqnarray}  
x(t)&=&[R+b\cos(6\pi t)]\sin(4\pi t)\nonumber\\
y(t)&=&[R+b\cos(6\pi t)] \cos(4\pi t) \nonumber\\
z(t)&=&b \sin(6\pi t)
\end{eqnarray}
where $R$ and $b$ are the major and minor axis of the torus, respectively, and $t \in [0,1]$ is the parameterization of the curve. Again, we define the tube by forming unit disks in the plane perpendicular to the curves tangent vector.  Pieranski and Przybyl \cite{trefoil} found that among these embedded trefoils, the minimum global curvature occurs at $R=1.1158$ and $b=1/2$.  As shown in Figure \ref{trefoilgraph}, our numerical integration of the excluded volume shows the same trend. As $R$ approaches $1.1158$, the excluded volume continuously decreases to its minimum at $R=1.1158$. Moreover, $R$ can be no smaller than $1.1158$ before the tube starts to self-overlap. Thus, as we argued, the depletion interaction drives the knot to its most compact structure as the depletors become small.

Note that the ideal trefoil knot ({\sl i.e.} the one that minimizes the global curvature) cannot be tied on a torus.  While there is no conformation proven to be the ideal trefoil knot, an upper bound on the minimum length of a trefoil knot has been found which cannot be achieved by a torus knot \cite{trefoil,idealtref}.    This was found via the ``shrink on no overlap'' (SONO) algorithm, which is similar in spirit to maximizing the overlap volume.  It is likely that these two approaches are equivalent as $r_s/t \rightarrow 0$ since our arguments related to global curvature should as well apply to the SONO algorithm.

\section{Beyond the Free Tube: Additional Interactions}

 \subsection{Tube Rigidity}
\begin{figure}[b]
\begin{center}
\includegraphics[width=3in]{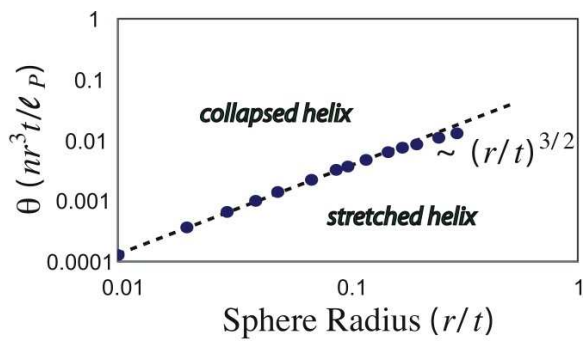} 
\caption{{The value of $\theta$ at which the gain in entropy of the spheres between a straight tube and the compact helix is equal to the bending energy of the helix, as a function of the sphere size.  When sphere concentration increases or the persistence length decreases the tube will go from a stretched helix to a collapsed (compact) helix.  For small sphere sizes the points roughly fall on the curve $(r_s/t)^{3/2}$, consistent with simple calculations done near the completely collapsed helix. }}
\label{bendingE}
\end{center}
\end{figure}

Real tubes, as opposed to the ideal tubes we have been considering, have a bending rigidity. The bending energy will compete with the entropy of the depleting spheres in selecting the ideal helix.  As is standard with rigid rods, we introduce a bending energy which is controlled by $\ell_p$, the persistence length \cite{RMP}, the length scale over which the tube can bend with an energy of order $k_BT$.  The bending energy can be written locally in terms of the curvature of the tube, $\kappa(s)$:
\begin{equation}
F_{elastic}= {1 \over 2} k_BT \ell_p\int \kappa^2(s) ds .
\end{equation}
This can be specialized to the helix for which $\kappa$ is independent of $s$ and equal to
$\kappa^{-1}=R\left[1+P^2/(2\pi R)^2\right]$.
Combining the bending energy with the entropy of the spheres allows us to calculate the change in free energy between a straight tube and a helix: 
 \begin{equation}\label{deltaF}
 \Delta F \approx -k_BTn V_{overlap}+{1 \over 2} {k_BT L\ell_p  R^2\over \left[R^2 + P^2/(2\pi)^2\right]^2} .
 \end{equation}
These two energies are minimized at opposite extremes. The bending energy is minimized for a straight tube with $R/P =0$, while the excluded volume free energy is minimized by the compact helix. To investigate this balance, we find the densities and persistence lengths for which $\Delta F=0$ between the straight tube and the compact helix for that sphere size.  The competition is controlled
by the relative size of $n$ and $\ell_p$ or, better, by the dimensionless quantities $nr_s^3$ and $\ell_p/t$, related to the volume fraction of the spheres and the aspect ratio of the tube, respectively.  We find it useful to define the control parameter $\theta\equiv nr_s^3/(\ell_p/t)$.  Again, for small $r_s/t$, we can estimate $V_{\rm overlap}/L \sim r_s^{3/2}t^{1/2}$ and, for the ideal compact helix, $\kappa\sim 1/t$.  Thus, we estimate the critical value of $\theta \sim (r_s/t)^{3/2}$, which is confirmed numerically by Figure \ref{bendingE}.  Note that as $r_s/t$ grows, the critical value of $\theta$ drops
below this value.  This follows because as $r_s/t$ grows, the optimal helix is not as tightly wound
and so the depletion effect need only overcome a smaller bending energy.

\begin{figure}
\begin{center}
\includegraphics[width=3in]{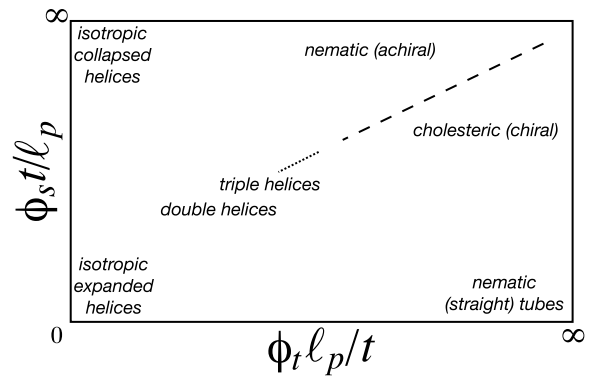} 
\caption{{Hypothetical phase diagram for a solution of tubes and spheres. }}
\label{phase}
\end{center}
\end{figure}

We identify the control parameter  as the ratio of {\sl two} dimensionless parameters, $\phi_s=4\pi nr_s^3/3$, the volume fraction of the depleting spheres and $\ell_p/t$, the parameter which controls the isotropic-to-nematic transition in the semiflexible tubes.  This suggests that we have just probed
one corner of a broader phase diagram where we must also include the volume fraction of the tubes, $\phi_t$.  Roughly speaking, when $\phi_t \gtrapprox 4t/\ell_p$, the tubes should form a nematic phase.  We sketch out a hypothetical phase diagram in Figure \ref{phase}.  It is an open and interesting question as to whether the tubes will form a nematic phase with equal numbers of left- and right-handed helices or, at a high enough concentration $\phi_t$, will form a 
cholesteric phase composed only of one type of helix.  We can start to delve into that question by calculating the total excluded volume of two helices as they approach each other.   We can get a sense of the difference in the overlap volume when two helices of opposite handedness are brought together versus two helices of the same handedness.  

   In the case of two helices there is no special symmetry to exploit and simplify our calculation of the excluded volume; we are now forced to do the full three dimensional numerical integration, a more computationally intensive endeavor.   In this more complicated situation there are many more degrees of freedom as well; the geometries of the respective helices, the distance between the helices, the relative orientation of the helices, and the relative phase difference between them at the point of contact can all alter the excluded volume.  For convenience we limit ourselves to the case where the two helices are both in the maximally compact configuration for isolated tubes and both have the same orientation so that overlap scales with length as opposed to contact at one point.  This still allows us to explore the dependence of the excluded volume on chirality and distance between helices.  All points in the first helix are defined as before by eq. \ref{helixDef}, and the points in a second helix of same handedness and no phase offset a distance $D$ apart are defined by
   \begin{equation} \label{helixDef2}
{\bf H_2}(\xi, \phi, s) = {\bf R}(\xi)+D\hat{x}+s \left[\cos{\phi}\; {\bf N}(\xi)+\sin{\phi}\;{\bf B}(\xi)\right]
\end{equation}
for $\phi\in[0,2\pi]$, $s\in[0,t]$, and $\xi\in[0,2\pi n]$, and {\bf N}, {\bf B}, and {\bf R} are defined by eq. \ref{helixvecs}.  In this equation the chirality of the second helix is changed by replacing $\xi$ by $-\xi$, and a phase offset of $\psi$ radians is added to the second helix by replacing $\xi$ by $\xi + \psi/(2\pi)$.  

\begin{figure} 
\begin{center}
\includegraphics[width=3in]{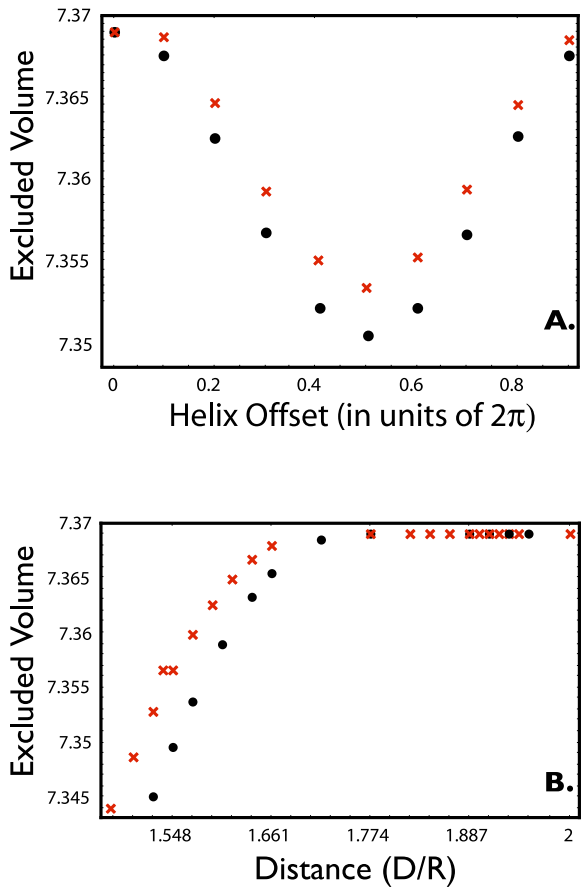} 
\caption{{Excluded volume for two maximally compact, aligned helices for sphere size of $r_s=0.1t$.  In A. the distance between the axis of the helix is held constant at $2R$ while the vertical offset is varied from 0 and $P$.  The circles are for two helices with the same handedness, while the {\sl x}'s are for two helices with opposite handedness.    For the same chirality the excluded volume is minimal when the helices are completly out of phase.   In B. the offset is set at 0 and the we bring the helices closer.  At any given distance, when the two helices have the same handedness there is less excluded volume, but the helices of opposite handedness are able to get closer, reducing the ultimate excluded volume. }}
\label{2helix}
\end{center}
\end{figure}    

    In Figure \ref{2helix}A, we can see the overlap volume between helices of the same handedness and opposite handedness whose centers are at a distance $2R$ apart.  The excluded volume is plotted as a function of the phase offset between the two helices, and the minimum occurs at an offset of $\pi$ when the two helices are just touching.  As the offset either increases or decreases the helices stop touching and the excluded volume increases.  From this graph one can see that for all offsets, two helices of the same chirality have less excluded volume then for two of opposite chirality.  This result is misleading on the surface. Unlike the case of one tube, the local density of the tubes contributes to the entropy balance. In this case there is more overlap between two helices of the same handedness because they can not pack as densely and start to touch at further distances than two helices of opposite chirality.  This is similar to the situation for hard spheres: the fcc lattice is the densest packing of spheres so that at a given density of spheres, the fcc lattice is the arrangement that gives spheres the most, not the least, free volume. Similarly helices of opposite handedness can move closer and will have more overlap volume at any given separation.  One should also note that even ignoring chirality, the excluded volume dependence on offset shown in Figure \ref{2helix}A appears strikingly similar to an anti-ferromagnetic XY model.  This has been shown in dense packing of long chains to have complex crystallization which are accompanied by spontaneous breaking of chiral symmetry \cite{grason}.
    
In Figure \ref{2helix}B we show the excluded volume as enantiomeric helices are brought closer than $2R$ apart.   Again, at every distance two helices of the same handedness have less excluded volume than two helices of opposite handedness, but eventually we get to the point where the two helices of equal handedness touch and can't move any closer.  The two helices of opposite handedness can still move slightly closer and when they nearly touch the excluded volume becomes lower than that for  two helices of the same chirality at close packing.  This shows that tubes of opposite handedness are favored as close-packed neighbors over two helices of the same chirality, but the difference is small.  
	
This brings us back to Figure \ref{phase} and our sketch of the phase diagram.  The results here imply that for small $\phi_t$ we will have an isotropic state.  As $\phi_t$ grows beyond $4t/\ell_p$, the tubes will align in a racemic state with roughly equal numbers of left- and right-handed helices.  As $\phi_t$ grows, for sufficiently large $\phi_s$, we expect a close-packed lattice of helices.  Since it is impossible to decorate the close-packed triangular lattice in an anti-chiral fashion where every pair of nearest neighbors is of opposite handedness, we expect that the aligned, racemic nematic will give way to a cholesteric phase of helices of uniform handedness.   In the middle of this phase diagram we can only speculate what would be seen -- perhaps    
more complex double and triple helix bundles might form as the tube concentration increases from the low concentration isotropic state. These complexes have been studied in the context of optimal geometry and so we should expect them to appear as tubes are brought together \cite{nplies}.  From there the large number of degrees of freedom creates a vast array of possibilities \cite{Saxe}.

\section{Conclusion}
We have shown how the depletion interaction can drive helix formation and have estimated at what lengths the simple helix should give way to more complex tube conformations.  In addition to providing a local, force-based method for calculating the minimally compact curve, our approach also allows for direct comparison to experimental
systems with, for instance, wormlike micelles in a solution of spherical micelles.  Our approach complements other purely geometrical approaches including the pioneering work of Maritan, Micheletti, Trovato, and Banavar \cite{Maritan}, the use of the global radius of curvature \cite{Maddocks}, and the use of overlap constraints \cite{Pieranski}.  The coincidence among the prior results and ours for the optimal helix strongly suggests that there is a primitive set of geometric motifs that are the building blocks of long, tube-like molecules \cite{Banavar}.  We have, in addition, found that maximizing free volume finds the optimal trefoil knot and promotes the formation of paired, opposite-handed helices. 
Our model has the added advantage of another, independent parameter, the sphere radius $r_s$.  Thus the depleting spheres can act as surrogates for other interactions such as hydrophobicity and polymer-polymer interactions, or even as the side chains to amino acids when considering protein conformations.
We have also considered the effects of tube stiffness and compared these to the purely entropic depletion forces.  Further work will more fully explore our proposed phase diagram (Figure \ref{phase}) and the higher order structures that are formed by the combination of nematic interactions and depletion forces.

\acknowledgments
It is a pleasure to acknowledge stimulating conversations with Carol Deutsch, Dennis Discher, Andrea Liu, and Arjun Yodh.  This work was supported by the University of Pennsylvania NSF MRSEC Grant DMR05-20020 and a gift from Lawrence J. Bernstein.  RDK thanks the Aspen Center for Physics where some of this work was performed.


\begin{thebibliography}{99}
\bibitem{Saxe} In guessing the possible phases we feel like the men of Indostan: ``And so these men of Indostan / Disputed loud and long, / Each in his own opinion/ Exceeding stiff and strong, / Though each was partly in the right, / And all were in the wrong!'', from Blind Men and the Elephant, by John Godfrey Saxe.
\end{thebibliography}
\end{document}